\documentclass[submission,copyright,creativecommons]{eptcs}
\providecommand{\event}{GandALF 2024} % Name of the event you are submitting to
\usepackage{iftex}
\ifpdf
  \usepackage{underscore}         % Only needed if you use pdflatex.
  \usepackage[T1]{fontenc}        % Recommended with pdflatex
\else
  \usepackage{breakurl}           % Not needed if you use pdflatex only.
\fi

\usepackage{listings}
\usepackage{subfigure}
\usepackage{xspace}
\usepackage{graphicx}
\usepackage[utf8]{inputenc}
\usepackage{multirow}
\usepackage{enumitem}
\usepackage{rotating}
\usepackage{float}

\usepackage{algpseudocode}
\usepackage{graphicx}
\usepackage{amsmath}
\usepackage{mathtools}
\usepackage{algorithm}
\usepackage{syntax}
\usepackage{amssymb}
\usepackage{array}
\newcommand{\bigB}[1]{\big[ #1 \big]}

\newcommand{\deq}{{\stackrel{\mathrm{def}}{=}}}
\newcommand{\defiff}{{\stackrel{\mathrm{def}}{\mIff}}}

\newcommand{\C}{{\mathcal{C}}}

\newcommand{\semiring}[5]{\big(#1, #2, #3, #4, #5\big)}
	\usepackage{stackrel}
\newcommand{\nAnd}{\;\mathrel{\wedge}\;}

\newcommand{\mImp}{\;\Longrightarrow\;}  
  
\newcommand{\mIff}{\;\Longleftrightarrow\;} 

\newcommand{\lnotation}[4]{
	\def\third:{#3} 
	\def\possiblyone:{} 
	\def\possiblytwo:{~}
	\def\possiblythree:{ }
	\def\divide{\;#1\hspace*{-0pt}( #2\; \mid: \; #4 \, )}
	\def\nodivide{\;#1\hspace*{-0pt}( #2\;\mid\; #3\;:\;#4 \, )}
	\ifx\third\possiblyone\divide
		\else\ifx\third\possiblytwo\divide
		\else \ifx\third\possiblythree\divide
		\else \nodivide\fi\fi\fi}

\newcommand{\biglnotation}[4]{
	\def\third:{#3} 
	\def\possiblyone:{} 
	\def\possiblytwo:{~}
	\def\possiblythree:{ }
	\def\divide{\;#1\hspace*{-0pt}\big( #2\; \mid: \; #4 \, \big)}
	\def\nodivide{\;#1\hspace*{-0pt}\big( #2\;\mid\; #3\;:\;#4 \, \big)}
	\ifx\third\possiblyone\divide
		\else\ifx\third\possiblytwo\divide
		\else \ifx\third\possiblythree\divide
		\else \nodivide\fi\fi\fi}

\newcommand{\bigglnotation}[4]{
	\def\third:{#3} 
	\def\possiblyone:{} 
	\def\possiblytwo:{~}
	\def\possiblythree:{ }
	\def\divide{\;#1\hspace*{-0pt}\bigg( #2\; \mid: \; #4 \, \bigg)}
	\def\nodivide{\;#1\hspace*{-0pt}\bigg( #2\;\mid\; #3\;:\;#4 \, \bigg)}
	\ifx\third\possiblyone\divide
		\else\ifx\third\possiblytwo\divide
		\else \ifx\third\possiblythree\divide
		\else \nodivide\fi\fi\fi}

\newcommand{\grieslnotation}[4]{
	\def\third:{#3} 
	\def\possiblyone:{} 
	\def\possiblytwo:{~}
	\def\possiblythree:{ }
	\def\divide{(#1 #2\; \mid : \; #4 \, )}
	\def\nodivide{(#1 #2\;\mid\; #3\;:\;#4 \, )}
	\ifx\third\possiblyone\divide
		\else\ifx\third\possiblytwo\divide
		\else \ifx\third\possiblythree\divide
		\else \nodivide\fi\fi\fi}
	\usepackage{color}
	\usepackage{etoolbox}
	\usepackage{soul}
\definecolor{darkred}{rgb}{0.75,0.0,0.0}
\definecolor{darkgreen}{rgb}{0.0,0.6,0.0}
\definecolor{darkblue}{rgb}{0.0,0.0,0.6}
\definecolor{darkcyan}{rgb}{0.0,0.6,0.6}
\definecolor{darkmagenta}{rgb}{0.6,0.0,0.6}
\definecolor{darkamber}{rgb}{1.0,0.5,0.0}
\definecolor{darkyellow}{rgb}{0.6,0.6,0.0}

\definecolor{lightred}{rgb}{1.0,0.9,0.9}
\definecolor{lightgreen}{rgb}{0.9,1.0,0.9}
\definecolor{lightblue}{rgb}{0.9,0.9,1.0}
\definecolor{lightcyan}{rgb}{0.8,1.0,1.0}
\definecolor{lightmagenta}{rgb}{1.0,0.8,1.0}
\definecolor{lightamber}{rgb}{1.0,0.8,0.0}
\definecolor{lightyellow}{rgb}{1.0,1.0,0.8}

\definecolor{webgreen}{rgb}{0,0.5,0}
\definecolor{webbrown}{rgb}{0.6,0,0}

\definecolor{grey}{rgb}{0.65,0.65,0.65}
\definecolor{purple}{rgb}{0.4,0,0.75}

\definecolor{burgundy}{rgb}{0.5, 0.0, 0.13}         % For states
\definecolor{darkcyan}{rgb}{0.0,0.6,0.6}            % For sync messages
\definecolor{darkpastelgreen}{rgb}{0.01, 0.75, 0.24}% For guarded transitions

\newcommand{\mynote}[2]{
	\ifstrequal{#1}{0}{\textcolor{darkamber}{#2}}{}%
  	\ifstrequal{#1}{1}{\textcolor{darkmagenta}{#2}}{}%
  	\ifstrequal{#1}{2}{\textcolor{darkcyan}{#2}}{}%
  	\ifstrequal{#1}{3}{\textcolor{darkgreen}{#2}}{}%
  	\ifstrequal{#1}{5}{\textcolor{darkblue}{#2}}{}%
  	\ifstrequal{#1}{8}{\textcolor{burgundy}{#2}}{}
}
\newcommand{\todiscuss}[2]{
	\ifstrequal{#1}{0}{\textcolor{darkamber}{\textit{\textbf{TO DISCUSS}: #2}}}{}%
  	\ifstrequal{#1}{1}{\textcolor{darkmagenta}{\textit{\textbf{TO DISCUSS}: #2}}}{}%
  	\ifstrequal{#1}{2}{\textcolor{darkcyan}{\textit{\textbf{TO DISCUSS}: #2}}}{}%
  	\ifstrequal{#1}{3}{\textcolor{darkgreen}{\textit{\textbf{TO DISCUSS}: #2}}}{}
}
\newcommand{\todo}[2]{
	\ifstrequal{#1}{0}{\textcolor{darkamber}{\textbf{\underline{TO DO}}: #2}}{}%
  	\ifstrequal{#1}{1}{\textcolor{darkmagenta}{\textbf{\underline{TO DO}}: #2}}{}%
  	\ifstrequal{#1}{2}{\textcolor{darkcyan}{\textbf{\underline{TO DO}}: #2}}{}%
  	\ifstrequal{#1}{3}{\textcolor{darkgreen}{\textbf{\underline{TO DO}}: #2}}{}
}
\newcommand{\toaddress}[2]{
	\ifstrequal{#1}{0}{\noindent\textcolor{darkamber}{$\bigstar$~\textbf{#2}}\\}{}%
  	\ifstrequal{#1}{1}{\noindent\textcolor{darkmagenta}{$\bigstar$~\textbf{#2}}\\}{}%
  	\ifstrequal{#1}{2}{\noindent\textcolor{darkcyan}{$\bigstar$~\textbf{#2}}\\}{}%
  	\ifstrequal{#1}{3}{\noindent\textcolor{darkgreen}{$\bigstar$~\textbf{#2}}\\}{}
}
\newcommand{\eg}{\textrm{e.g.,}\@\xspace}
\newcommand{\ie}{\textrm{i.e.,}\@\xspace}

\newcommand{\etal}{\textrm{et~al.}\@\xspace}

\newcommand{\opt}[1]{\mathit{opt}\bigB{#1}}
\newcommand{\refines}{\sqsubseteq}
\newcommand{\requires}[3]{#1 \xrightarrow[]{#3} #2}
\newcommand{\firebird}{\textsl{Firebird}\@\xspace}

\newtheorem{definition}{Definition}

\newlist{tabitemize}{itemize}{1}
\setlist[tabitemize]{label=\textbullet, 
                     leftmargin=*,
                     nosep, 
                     before=\begin{minipage}[t]{\hsize}\raggedright, 
                     after=\end{minipage}}
\title{A Game-Theoretic Approach for Security Control Selection\thanks{Funded in part by the Human-Centric Cybersecurity Partnership under the SSHRC Partnership Grants program.}}
\author{
Dylan L\'{e}veill\'{e} \qquad\quad\qquad Jason Jaskolka
\institute{Department of Systems and Computer Engineering\\
Carleton University, Ottawa, ON, Canada}
\email{\quad dylan.leveille@carleton.ca \quad jason.jaskolka@carleton.ca}
}
\def\titlerunning{A Game-Theoretic Approach to Security Control Selection}
\def\authorrunning{D. L\'{e}veill\'{e} \& J. Jaskolka}
\begin{document}
\maketitle

\begin{abstract}
    
Selecting the combination of security controls that will most effectively protect a system's assets is a difficult task. If the wrong controls are selected, the system may be left vulnerable to cyber-attacks that can impact the confidentiality, integrity and availability of critical data and services. In practical settings, it is not possible to select and implement every control possible. Instead considerations, such as budget, effectiveness, and dependencies among various controls, must be considered to choose a combination of security controls that best achieve a set of system security objectives. In this paper, we propose a game-theoretic approach for selecting effective combinations of security controls based on expected attacker profiles and a set budget. The control selection problem is set up as a two-person zero-sum one-shot game. Valid control combinations for selection are generated using an algebraic formalism to account for dependencies among selected controls. We demonstrate the proposed approach on an illustrative financial system used in government departments under four different scenarios. The results illustrate how a security analyst can use the proposed approach to guide and support decision-making in the  control selection activity when developing secure systems.
\end{abstract}

% Paper Body
\section{Introduction}
\label{sec:introduction}
% Begin Section

With computers becoming more interconnected than ever, there emerges an even greater need to secure computer systems and to effectively manage security risks. Security risks are mitigated by the implementation of a set of security controls. A \textit{security control} refers to a safeguard or countermeasure prescribed for an information system or an organization designed to protect the confidentiality, integrity, and availability of its information and to meet a set of defined security requirements~\cite{NIST-800-160v2r1}.  

\textit{Control selection} is an activity commonly found as part of a risk management process~\cite{ISO-31000}, a systems engineering process~\cite{NIST-800-160v2r1}, the Risk Management Framework~\cite{NIST-RMF}, the Cybersecurity Framework~\cite{NIST-CSF}, or the Privacy Framework~\cite{NIST-PF}.
Control selection involves selecting and documenting the security controls necessary to protect the information system and organization commensurate with risk to organizational and system operations and assets, individuals, other organizations, and the nation~\cite{NIST-RMF}. 

During the control selection activity, security analysts typically select security controls from standardized security control catalogues, such as NIST~SP~800-53~\cite{NIST-800-53r5}, ITSG-33~\cite{ITSG-33}, ISO~27002~\cite{ISO-27002}, CIS Critical Security Controls~\cite{cisControls}, and MITRE D3FEND\textsuperscript{\texttrademark}~\cite{DEFEND}, among others. However, selecting combinations of controls from these catalogues can be difficult for several reasons. 
First, these control catalogues are large, and many possible controls could be selected to mitigate the risks identified for a given system. In practical settings, it is not possible to select and implement every control possible. Considerations such as budget, effectiveness, and dependencies among various controls, must be considered to choose a combination of security controls that best achieve a set of system security objectives. 
Second, control selection is largely a human-oriented activity. The dynamics between security analysts (defenders) strategizing to protect critical systems and assets and achieve a set of security objectives, and attackers aiming to impact critical systems and assets and violate those same security objectives must be considered when deciding on the most effective and cost-efficient combination of security controls. 
Although numerous optimization-based solutions are adept at accounting for various properties of the controls themselves, they fail to capture the human element that is inherently part of the control selection activity. 

To address the above mentioned challenges, we propose a game-theoretic approach for security control selection. The human aspects of the control selection problem, as well as the large space of possible control combinations, and their dependencies and constraints, lends itself well to an application of game theory. Specifically, we set up a two-person zero-sum one-shot game which is played by a security analyst. The analyst selects their strategy based on an attacker profile, characterized by the expected targeted assets and security objectives. Each analyst strategy corresponds to a combination of security controls from a chosen control catalogue that are capable of achieving the security objectives. Valid control combinations are generated using an algebraic formalism (akin to product family algebra~\cite{Hofner2011}) to account for dependencies among selected controls. The outcome of the game is a combination of suggested security controls that can effectively defend against the considered attacker profile. Using an illustrative governmental finance system, we demonstrate the proposed approach under four different scenarios. 

The rest of this paper is organized as follows. Section~\ref{sec:relatedwork} provides an overview of existing works on the topic of control selection and of game theory applications in cybersecurity. Section~\ref{sec:approach} presents the proposed game-theoretic approach for control selection. Section~\ref{sec:example} provides an illustrative example demonstrating the application of the proposed approach. Section~\ref{sec:discussion} discusses the benefits and potential difficulties with the proposed approach. Lastly, Section~\ref{sec:conclusion} concludes and briefly discusses future work.
% End Section

\section{Related Work}
\label{sec:relatedwork}
% Begin Section
Many existing approaches to support the security control selection activity are based on setting and solving optimization problems. For example, for each considered control, Yevseyeva \etal~\cite{Yevseyeva2015} assign a probability of ``survival'' for each possible threat (\ie the probability that the threat persists in the presence of the control). Probabilities are also assigned for the expected loss of successful attacks. The goal of the proposed approach is to minimize this expected loss, under constraints such as cost and system resources. Similarly, Almeida and Respício~\cite{Almeida2018} also assign probabilities to controls based on their expected performance in mitigating certain vulnerabilities. For the proposed approach, the goal is to find the optimal controls for the system that will minimize an objective function accounting for both loss and cost. A different approach was proposed by Dewri \etal~\cite{Dewri2007} where systems are modelled as trees, in which the leaf nodes represent possible attacks. Controls therefore mitigate one or many leaf nodes. With the attack impact, attack frequency, and cost of each control known, the optimal controls can be found by optimization. A similar tree-like approach was also proposed by Park and Huh~\cite{Park2020}. While optimization-based approaches can account for important considerations and constraints such as cost and effectiveness, they depend heavily on assumptions about probabilities for threat likelihoods, or control success rates. Such probabilities are not likely to be accurately known in a practical setting. 

Several other approaches for control selection that are not based on optimization have also been proposed.
Bettaieb \etal~\cite{Bettaieb2020} presented an approach where a machine learning model is trained with historical data from previous security assessments to make predictions using certain features of interest from a given security assessment to determine optimal controls. However, using historic data to determine how to protect a system has several limitations as every system is unique and may operate in widely different environments. In another work, Kiesling \etal~\cite{Kiesling2016} proposed a simulation-based approach to determine the optimal controls for a system. To do this, expected attacks are simulated on different components of the system using different possible control combinations to find the optimal ones. This approach is noteworthy as it simply uses the properties of the controls and of the current system (such as different threats) to find the most optimal control combinations and does not depend on any probabilities.

Several works have explored the use of game theory for addressing cybersecurity challenges. 
For example, Nassar \etal~\cite{Nassar2021} proposed a technique which focused on evaluating a system's network security with the help of a game model. Smith \etal~\cite{Smith2017} used game theory to verify the security of hardware designs. Wang \etal~\cite{Wang2010} presented a network attack-defence game to help secure a computer network. However, game theory has yet to be utilized for security control selection.

In contrast to existing work, the proposed approach aims to leverage game theory to address the shortcomings of current control selection approaches by placing a central focus on possible attacker behaviours, while also taking into account the considerations and constraints that limit the selection of certain combinations of security controls to effectively mitigate the threats to a system. 
% End Section

\section{The Proposed Approach}
\label{sec:approach}
\begin{figure}[ht!]
    \centering
    \includegraphics[width=\textwidth]{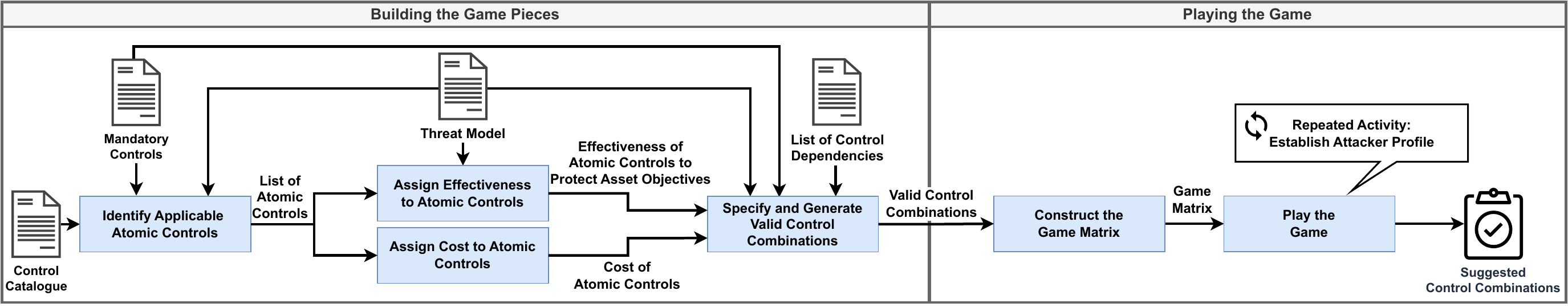}
    \vspace{-2em}
    \caption{An overview of the proposed game-theoretic approach for security control selection}
    \label{fig:propMeth}
\end{figure}

% Begin Section
In this section, we present our proposed game-theoretic approach for security control selection. An overview of the approach is shown in Figure~\ref{fig:propMeth}.
The approach consists of two main stages shown as swim lanes and six steps shown in blue. 
All steps are to be conducted by a security analyst. A detailed description of each step of the proposed approach is provided in the sections below. 

\subsection{Identify Applicable Atomic Controls}

The approach starts with the security analyst identifying applicable atomic controls from a given security control catalogue. In our context, the \textit{atomic controls} are the smallest (indivisible) security controls that can be selected from a control catalogue. We say that a control is \textit{applicable} to a system if it could provide any form of protection from the threats to the assets of the system. 

We assume that a list of threats and assets are available to the security analyst in the form of a threat model. A threat model is defined as ``a structured representation of all the information that affects the security of an application''~\cite{Drake}. Threat models typically include identified system threats and their impact on the assets within the system~\cite{Drake,NIST-800-154}. 
The threat model can be obtained by applying a well-known threat modelling methodology such as STRIDE~\cite{STRIDE} or PASTA~\cite{PASTA}.

To determine the applicability of an atomic control, the security analyst must carefully consider each atomic control from the given control catalogue and decide if the control can mitigate the identified threats to the assets. Additionally, certain organizational needs or standards and regulations for the system's application domain may require that specific security controls be present in the system. The security analyst must therefore ensure that these \textit{mandatory controls} are included as part of the set of applicable controls identified. 
This is a manual process. However, it should be noted that the effort required for this activity is reasonable as security control catalogues are typically separated by control families which help guide an analyst in finding suitable controls~\cite{Rouland2023ab}. Instructions and guidance for performing this task is well documented by ISO~27005~\cite{ISO-27005} and NIST~SP~800-53B~\cite{NIST-800-53B}.

\textit{At the end of this step, the analyst will have a set of applicable atomic controls for the system.} Note that the combination of suggested controls found by applying the proposed approach will be a subset of the controls gathered in this initial step.

\subsection{Assign Effectiveness to Atomic Controls}
\label{sec:approach:eff}

For each identified atomic control, the analyst proceeds by assigning an effectiveness of the control at satisfying each \textit{security objective} on each asset in the system. Security objectives represent the security needs of the assets on the system, such as confidentiality, integrity, and availability~\cite{NIST-1800-26}. These objectives are normally included as part of the threat impacts described in the threat model. It is important to remember that the goal of the proposed approach is to create a game. In every game, there needs to be strategies, and payoffs defined for each strategy. Assigning the effectiveness of each atomic control therefore defines the payoffs of each atomic control in the game. 

To perform this step of the approach, the atomic payoff matrix presented in Table~\ref{tab:atomicPayoff} must be completed. The rows represent each atomic control that was identified in the previous step (denoted $C_{1},\dots,C_{N}$). The columns represent the security objectives for each asset (denoted $O_{1},\dots,O_{M}$). We expect the analyst to assign a value between 0 and 1 in each cell of this matrix. A value of 0 means that the atomic control is not effective at satisfying the specified objective for an asset, while a value of 1 means that the atomic control is completely effective at satisfying the specified objective for an asset. Each payoff value is therefore normalized. Provided that the rating scheme is selected and used consistently throughout the approach, the analyst is free to choose any method for assigning the effectiveness values for the atomic payoff matrix. For example, the analyst may choose to use a quantitative approach as in the Defect Detection and Prevention (DDP) risk reduction strategy developed by NASA~\cite{Feather2005}, or they may alternatively choose to use a qualitative rating mapped to quantitative values as in~\cite{Liu2011aa}.

\begin{table}[ht!]
\caption{General form of the atomic payoff matrix}
\label{tab:atomicPayoff}
    \centering
    \small
    \begin{tabular}{|c|c|c|c|c|c|c|c|c|c|c|c|c|c|c|}
    \cline{2-11}
    \multicolumn{1}{c|}{} & \multicolumn{3}{c|}{\textit{Asset 1}} & \multicolumn{3}{c|}{\textit{Asset 2}} & \multicolumn{1}{c|}{\dots} & \multicolumn{3}{c|}{\textit{Asset X}}\\
    %  \cline{6-8}
    %   \multicolumn{5}{c|}{} & \multicolumn{5}{c|}{numbers2} \\
    \cline{2-11}
    \multicolumn{1}{c|}{} & $O_{1}$ & \dots & $O_{M}$ & $O_{1}$ & \dots & $O_{M}$ & \dots & $O_{1}$ & \dots & $O_{M}$ \\
    \hline
    $C_{1}$ &  &  &  & & & & &  & &\\
    \cline{1-11}
    % $C_{2}$ &  &  &  & & & & &  & &\\
    % \cline{1-11}
    \vdots &  &  &  & & & & &  & &\\
    \cline{1-11}
    $C_{N}$ &  &  &  & & & & &  & &\\
    \hline
    \end{tabular}
\end{table}

\textit{At the end of this step, the analyst will have the effectiveness of each applicable atomic control for satisfying each security objective on each asset in the system. }

\subsection{Assign Cost to Atomic Controls}
\label{sec:approach:cost}

In practical settings, cost or time constraints limit how many controls can be part of a system; if there are too many controls they may exceed a certain budget or cannot be implemented in reasonable time. In fact, without such constraints, there could technically be no limitations on the number of controls that can be selected for a system, and the best solution would be to select them all. 

At the same time as assigning effectiveness, the analyst will also need to assign a cost for each identified atomic control. We expect the analyst to assign a cost from the set of real numbers $\mathbb{R}$. The units for cost could be represented as dollars, thousands of dollars, or any other form of currency as long as the same units are consistently used for all cost values. Furthermore, no units could be used if desired. Without units, costs simply represent an implementation effort.

\textit{After this step, the analyst will have the cost associated with each applicable atomic control. }

\subsection{Specify and Generate Valid Control Combinations}

Given a set of applicable atomic controls, the analyst needs to specify and generate the set of valid control combinations that satisfies their constraints. To formally capture these constraints, we have decided to use an algebraic specification based on product family algebra~\cite{Hofner2011} to specify and generate valid combinations of security controls.

Product family algebra extends the mathematical notions of semirings to describe and manipulate product families. A semiring is an algebraic structure~$\semiring{S}{+}{\cdot}{0}{1}$ consisting of a set $S$ with a commutative and associative binary operator $+$ and an associative binary operator $\cdot$. An element $0 \in S$ is the identity element with respect to $+$, while an element $1 \in S$ is the identity element with respect to $\cdot$. Additionally, $\cdot$ distributes over $+$ and element $0$ annihilates $S$ with respect to $\cdot$. A semiring is commutative if $\cdot$ is commutative and a semiring is idempotent if $+$ is idempotent.

 For ease of presentation, we recast the vocabulary of product family engineering into the vocabulary of security controls by first defining a security control algebra to express families of security control combinations generated from a set of atomic controls. 

\begin{definition}[Security Control Algebra]
\label{def:controlAlgebra}
    A \emph{security control algebra} is a commutative idempotent semiring~$\C\ \deq \semiring{C}{\oplus}{\odot}{0}{1}$ where each element of the semiring $c \in C$ is a security control family.
\end{definition}
In a security control algebra, the operator $\oplus$ is interpreted as a choice between two security control families and the operator $\odot$ is interpreted as a mandatory composition of two security control families\footnote{When the context is clear, we omit the mandatory composition operator $\odot$ when specifying security control algebra terms.}. The element $0$ represents a non-implementable security control combination that cannot exist and the element $1$ represents the empty security control combination which has no controls.
A security control family is called a \textit{security control combination} if it is indivisible with regard to the choice operator $\oplus$. Additionally, it is called a \textit{proper security control combination} if  $c \neq 0$. A security control combination is an \textit{atomic control} if is it is indivisible with regard to the mandatory composition operator~$\odot$.
Optional controls are expressed as a choice between the controls and the empty security control combination $1$. A list of optional controls $c_1,\dots,c_n$ is denoted by $\opt{c_1,\dots,c_n} \deq (c_1 \oplus 1) \odot \dots \odot (c_n \oplus 1)$. 

For two security control families $c_1$ and $c_2$ in a security control algebra, the \textit{refinement relation} ($\refines$) is defined as $c_1 \refines c_2 \defiff \lnotation{\exists}{c_3}{}{c_1 \leq c_2 \odot c_3}$ where $\leq$ is the natural semiring order (\ie $c_1 \leq c_2 \defiff c_1 \oplus c_2 = c_2$). 
To specify constraints, such as dependencies between controls, we use the requirement relation. 

\begin{definition}[Requirement Relation~\cite{Hofner2011}]
\label{def:reqRelation}
    For elements $c_1, c_2, c_3, c_4$ and security control combination $x$ in a security control algebra, the \textit{requirement relation} ($\rightarrow$) is defined inductively as:
    \begin{eqnarray*}
        \requires{c_1}{c_2}{x} &\defiff& x \refines c_1 \mImp x \refines c_2 \\
        \requires{c_1}{c_2}{c_3 \oplus c_4} &\defiff& \requires{c_1}{c_2}{c_3} \nAnd \requires{c_1}{c_2}{c_4}
    \end{eqnarray*}
\end{definition}
For elements $c_1, c_2$ and $x$, the requirement relation $\requires{c_1}{c_2}{x}$ can be read as ``$c_1$ requires $c_2$ within $x$.''

With this setting, all security control combinations can be specified algebraically by expressing the mandatory and optional controls as terms of a security control algebra along with requirement relations describing control dependencies.

The resulting specification serves as the basis for generating all possible proper security control combinations. However, not all control combinations are possible as some may exceed our defined budget. To make this determination we first define how to calculate the cost of a proper security control combination. In what follows, let $P \subseteq C$ be the set of all proper security control combinations in a security control algebra $\C$.

\begin{definition}[Cost of a Proper Security Control Combination]
\label{def:costInduction}
  The cost of a proper security control combination $\textit{Cost}: P \rightarrow \mathbb{R}$ is a function defined inductively for any proper security control combinations $a, b \in P$ in a security control algebra $\C$ as:
    \begin{eqnarray*}
        \mathit{Cost}(1) &=& 0\\
        \mathit{Cost}(a) &=& G(a)\ \text{if $a$ is atomic} \\
        \mathit{Cost}(a \odot b) &=& \mathit{Cost}(a) + \mathit{Cost}(b)
    \end{eqnarray*}
    where $G$ is a function that returns the cost assigned to an atomic control (see Section~\ref{sec:approach:cost}).    
\end{definition}

Now that we can compute the cost of a proper security control combination, we determine the set of valid security control combinations. A \textit{valid security control combination} is a proper security control combination that does not exceed the prescribed cost budget. The validity of a control combination is formalized in the following rule.
\begin{definition}[Budget Rule]
\label{def:rule}
    For any $p \in P$ and budget $B$:
    \begin{equation*}
        \mathit{Valid}(p) \mIff \mathit{Cost}(p) \leq B
    \end{equation*}
\end{definition}

\textit{After this step, the analyst will have a set of valid security control combinations that satisfy the prescribed budget.} These valid security control combinations become the strategies that an analyst can select when playing the game.

\subsection{Construct the Game Matrix}

In this step, the analyst constructs the game matrix. The general form of the game matrix can be seen in Table~\ref{tab:basicGameMatrix}. The rows represent the valid security control combinations found from the last step (denoted $\mathit{Combo}_{1},\dots,\mathit{Combo}_{N}$). The columns represent the security objectives for each asset (denoted $O_{1},\dots,O_{M}$). Note that the game matrix is identical in style to that of the atomic payoff matrix (see Table~\ref{tab:atomicPayoff}). The game matrix simply has control combinations as rows rather than atomic controls. 
In the game, the strategies of the security analyst will be the valid security control combinations, while the strategies of the attacker will be each security objective that could be violated on every asset. 

\begin{table}[ht!]
\caption{General form of the game matrix}
\label{tab:basicGameMatrix}
    \centering
    \small
    \begin{tabular}{l|lll|lll|l|lll|}
    \cline{2-11}
                             & \multicolumn{3}{c|}{\textit{Asset 1}}                            & \multicolumn{3}{c|}{\textit{Asset 2}}                            & \multicolumn{1}{c|}{\dots} & \multicolumn{3}{c|}{\textit{Asset X}}                            \\ \cline{2-11} 
                             & \multicolumn{1}{l|}{$O_{1}$} & \multicolumn{1}{l|}{\dots} & $O_{M}$ & \multicolumn{1}{l|}{$O_{1}$} & \multicolumn{1}{l|}{\dots} & $O_{M}$ & \dots                     & \multicolumn{1}{l|}{$O_{1}$} & \multicolumn{1}{l|}{\dots} & $O_{M}$ \\ \hline
    \multicolumn{1}{|l|}{$\mathit{Combo}_1$} & \multicolumn{1}{l|}{}   & \multicolumn{1}{l|}{}    &    & \multicolumn{1}{l|}{}   & \multicolumn{1}{l|}{}    &    &                          & \multicolumn{1}{l|}{}   & \multicolumn{1}{l|}{}    &    \\ \hline
    % \multicolumn{1}{|l|}{$\mathit{Combo}_2$} & \multicolumn{1}{l|}{}   & \multicolumn{1}{l|}{}    &    & \multicolumn{1}{l|}{}   & \multicolumn{1}{l|}{}    &    &                          & \multicolumn{1}{l|}{}   & \multicolumn{1}{l|}{}    &    \\ \hline
    \multicolumn{1}{|c|}{\vdots}   & \multicolumn{1}{l|}{}   & \multicolumn{1}{l|}{}    &    & \multicolumn{1}{l|}{}   & \multicolumn{1}{l|}{}    &    &                          & \multicolumn{1}{l|}{}   & \multicolumn{1}{l|}{}    &    \\ \hline
    \multicolumn{1}{|l|}{$\mathit{Combo}_N$} & \multicolumn{1}{l|}{}   & \multicolumn{1}{l|}{}    &    & \multicolumn{1}{l|}{}   & \multicolumn{1}{l|}{}    &    &                          & \multicolumn{1}{l|}{}   & \multicolumn{1}{l|}{}    &    \\ \hline
    \end{tabular}
\end{table}

Each outcome in a game is tied to a payoff~\cite{Straffin1993}. In our game, the payoffs are represented from the perspective of the analyst and represent the effectiveness of the security control combinations towards every asset's security objectives. Just as cost was defined inductively, we can define a proper control combination’s effectiveness towards an asset's security objective in a similar manner.

\begin{definition}[Effectiveness of a Proper Security Control Combination]
\label{def:effInduction}

    The effectiveness of a proper security control combination towards an asset's security objective $\mathit{Eff}: P \rightarrow \mathbb{R}$ is a function defined inductively for any proper security control combinations $a, b \in P$ in a security control algebra $\C$ as:
    \begin{eqnarray*}
        \mathit{Eff}(1) &=& 0\\
        \mathit{Eff}(a) &=& E(a)\ \text{if $a$ is atomic} \\
        \mathit{Eff}(a \odot b) &=& 1 - (1 - \mathit{Eff}(a))( 1- \mathit{Eff}(b) )
    \end{eqnarray*}
    where $E$ is a function that returns the effectiveness assigned to an atomic control for an asset's security objective (see Section~\ref{sec:approach:eff}). 
\end{definition}

With Definition~\ref{def:effInduction}, the payoff values in the game matrix can be calculated. Note that the calculation of the effectiveness of a security control combination is inspired from the combined effectiveness calculation as part of NASA's DDP approach~\cite{Feather2005}.

\textit{After this step, the analyst will have the game matrix so that they can proceed to play the game.}  

\subsection{Play the Game}

The game is a \textit{two-person zero-sum one-shot game}. The game is played by \textit{two persons}: the security analyst and the attacker. The attacker may embody one or multiple entities, but acts as a unified adversary. The goal of the security analyst is to select the security control combination that will best protect the security objectives for the assets they believe will be targeted by the attacker. Only one security control combination can be selected, hence it is a \textit{one-shot} game. On the other hand, the goal of the attacker is to attack assets and violate corresponding security objectives. 
An attacker could attack one or many assets and violate one or more objectives from a series of attacks. Regardless, an attacker will select which assets and objectives they will target and will commit to attacking the selected assets and objectives.
The attacker will naturally prefer attacking assets which are not properly defended, \ie those for which there are minimally effective security controls. The effectiveness values in the game matrix (payoffs) do not directly correlate to a loss to the attacker. However, it is easy to see that the higher the values, the more difficult it is for an attacker to conduct a successful attack leading to corresponding security objective violations. Therefore, what the security analyst gains in effectiveness is what the attacker loses in their ability to successfully conduct their attack; hence, it is a \textit{zero-sum} game. Note that this game is strictly non-cooperative; the analyst and attacker are competing directly and would never want to cooperate.

Using the game matrix, the analyst must select a strategy (\ie a valid security control combination) to play that  will best protect the system assets and security objectives that they believe are most important. To do this, an analyst must establish the expected attacker profile. An \textit{attacker profile} is an expected set of the assets and corresponding security objectives targeted by the attacker. One can imagine different classes of attackers having different capabilities, and different targets, thereby establishing different attacker profiles. In the context of a game, an attacker profile corresponds to guessing the attacker strategy so that it can be defended. This consideration of the dynamics of the analyst and the attacker strategies in this game is what differentiates it from existing security control selection approaches.  

It is impossible to know exactly which security objectives on which assets will be attacked, so assumptions must be made. One way to do this is to determine where most of the critical information flows in the system and which assets may be prone to more attacks (\ie have more expected threats). The combination of these ideas can help localize assets that are more attractive for attacks, and therefore puts the security objectives of these assets at higher risk of violation. Another way to do this is to consider the risk to each asset and corresponding security objectives for the identified threats to the system (which we consider known to the analyst). In this case, prioritizing defence of assets and security objectives targeted by  high risk threats may be a good approach. Regardless, once the attacker profile is determined, then the suggested strategy (\ie the most effective security control combination) can be found.

\begin{figure}[ht!]
    \centering
    \centerline{\includegraphics[width=\textwidth]{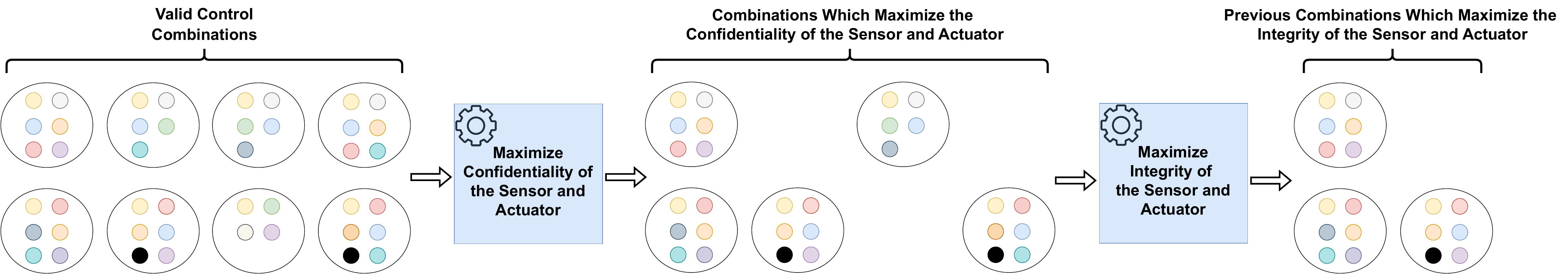}}
    \caption{Finding the suggested controls for an attacker profile with multiple ordered attacker objectives}
    \label{fig:combinedPersp}
\end{figure}

Regardless of the approach taken to establish the attacker profile, it will articulate the objectives that are expected to be violated by an attacker. For this work, we establish an attacker profile by considering and prioritizing different attacker objectives. Attacker objectives correspond to a set of security objectives for some assets that are equally expected to be targeted by an attacker. Within an attacker profile, several attacker objectives may be prioritized according to their perceived likelihood of being targeted by the attacker to obtain a priority order for the objectives. For example, the security analyst could establish an attacker profile in which the attacker has two ordered attacker objectives: (1) to target the confidentiality of two specific assets equally, and (2) to target the integrity of two other assets equally. The security analyst may consider as many attacker objectives as they desire when developing an attacker profile. The suggested analyst strategies for an attacker profile will be those which maximize the \textit{total effectiveness} across each attacker objectives (\ie the sum of the effectiveness returned by Definition~\ref{def:effInduction} for the security objectives in the attacker objectives is maximized in the priority order). To better understand this concept, an example of the strategies found by playing the game with an attacker profile with two ordered attacker objectives is visualized in Figure~\ref{fig:combinedPersp}.

In this example, there are initially eight valid security control combinations. Each security control combination has a unique set of controls (denoted by the different coloured dots in the figure). Only two  assets exist in this system; a Sensor and an Actuator. The attacker profile has two ordered attacker objectives: (1) the confidentiality of the Sensor and the Actuator and then (2) the integrity of the Sensor and the Actuator. From all valid control combinations, the control combinations which maximize the first set of attacker objectives is found, yielding five different combinations. From these five combinations, the combinations which maximize the second set of attacker objectives is found, yielding three control combinations. As there are no more ordered attacker objectives, the resulting control combinations are all considered equally valid, and represent the suggested strategies. Note that since the suggested strategies are derived through a series of maximization problems, it may be possible for more than one strategy to be the most effective for a given attacker profile.

\textit{At the end of this step, the analyst will obtain at least one strategy that best protects against the considered attacker profile and that corresponds to the suggested security control combinations to be implemented in the system.}  It is important to remember that this approach is a game. Therefore, as with any game, it is recommended that the game be re-constructed with different maximum budget values and re-played with different attacker profiles (as illustrated in Figure~\ref{fig:propMeth}). This can help gauge and compare the control combinations that should be used for the system under different constraints and goals.

% End Section

\section{Illustrative Example}
\label{sec:example}
% Begin Section

\begin{figure}[ht!]
    \centering
    \includegraphics[width=0.75\textwidth]{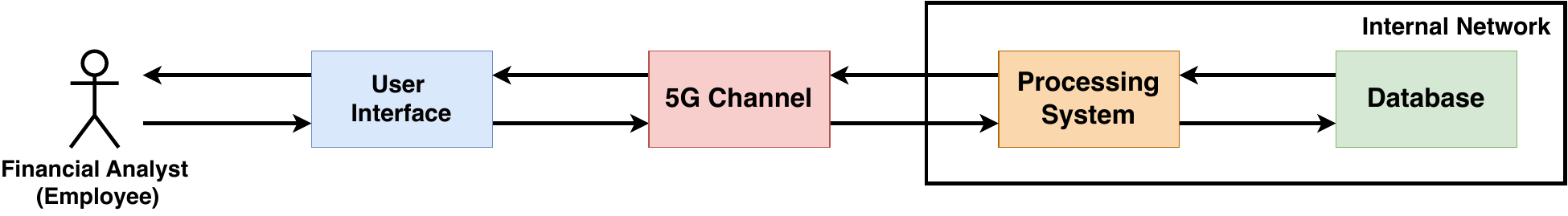}
    \vspace{-1em}
    \caption{An overview of the \firebird system architecture}
    \label{fig:toyArch}
\end{figure}

In this section, we demonstrate how the approach presented in Section~\ref{sec:approach} could be applied to support the control selection activity for an illustrative example system. Suppose a security analyst needs to select a combination of cost-effective security controls to protect a financial system used by the Canadian government called \firebird. An overview of the system architecture is shown in Figure~\ref{fig:toyArch}. \firebird allows financial analysts to enter data about financial transactions and view those transactions through a user interface. Many identical interfaces may exist. The interfaces communicate over a 5G channel to a central processing system to process the commands from the analyst. A database stores the financial transactions data used by the central processing system. Both the processing system and database are located in an internal government network. 
The security analyst will apply the proposed game-theoretic approach for security control selection for the \firebird system as described in the following sections. 
\vspace{-1.5em}

\subsection{Identify Applicable Atomic Controls}

Considering the \firebird system architecture shown in Figure~\ref{fig:toyArch}, there are four primary assets: the user interface, the 5G channel, the processing system, and the database. For simplicity and brevity, suppose that the analyst is primarily focused on addressing threats to the user interface and the database and threats to the 5G channel and processing system are being handled by another analyst. Also, it has been pre-determined by the security analyst's government department that they are primarily concerned with the confidentiality (C), integrity (I), and availability (A) security objectives for the system assets. 

Because \firebird is a Canadian government system, the analyst selects controls from the ITSG-33 control catalogue\footnote{ITSG-33 is the standard control catalogue to assist security practitioners in their efforts to protect information systems in compliance with applicable Government of Canada legislation, policies, directives, and standards~\cite{ITSG-33}.}. To comply with departmental requirements, it was decided by the security analyst's government department that the input validation control (\ie \textit{SI-10} in ITSG-33) must be present in the system. This is because improper input validation in any system can result in potentially severe consequences~\cite{khalaf2021web,scholte2012, Theodoor2012}.

The analyst is provided with the fragment of the threat model consisting of the assets, threats, and violated security objectives for the system as shown the first three columns of Table~\ref{tab:threatmodel}. 
By consulting the control catalogue, the analyst decides which of the atomic controls are relevant in protecting the system by referring to the identified threats in the threat model. 
The applicable atomic controls found for each threat can also be seen as part of Table~\ref{tab:threatmodel}. Notice that the mandatory control (\textit{SI-10:~Input Validation}) is included in the gathered set of atomic controls; the other controls therefore represent optional controls that may or may not be included as part of the suggested controls of this approach.

\vspace{-1em}
\begin{table}[ht!]
\caption{Threat model and applicable atomic controls for \firebird}
\label{tab:threatmodel}
\centering

\resizebox{\textwidth}{!}{
\begin{tabular}{|>{\centering\arraybackslash}p{0.6in}|>{\centering\arraybackslash}p{2.5in}|>{\centering\arraybackslash}p{1.3in}|>{\centering\arraybackslash}p{2.45in}|}
\hline
\textbf{Assets} & \textbf{Threats} & \textbf{Security Objectives Violated} & \textbf{Applicable Atomic Controls} \\ \hline
\multirow{3}{*}{\shortstack{User\\Interface}}       & 
\begin{tabitemize}
\item Commands received from unknown sources 
\end{tabitemize}
& 
\begin{tabitemize}
\item Confidentiality
\item Integrity
\end{tabitemize}
& 
\begin{tabitemize}
\item \textit{AC-4: Information Flow Enforcement}
\end{tabitemize}
\\ \cline{2-4} 
                              & 
\begin{tabitemize}
\item Improper/malicious commands entered  
\end{tabitemize}
& 
\begin{tabitemize}
\item Confidentiality
\item Integrity
\end{tabitemize}
& 
\begin{tabitemize}
\item \textit{SI-10: Input Validation}  
\end{tabitemize}
\\ \cline{2-4} 
                              & 
\begin{tabitemize}
\item Employee freely accesses and changes features provided in the interface 
\end{tabitemize}
& 
\begin{tabitemize}
\item Confidentiality
\item Integrity
\end{tabitemize}
& 
\begin{tabitemize}
\item \textit{AC-3: Access Enforcement}
\item \textit{AC-6: Least Privilege} 
\end{tabitemize}
\\ \hline
\multirow{2}{*}{Database}             & 
\begin{tabitemize}
\item SQL injection from an improper analyst input changes or retrieves data  
\end{tabitemize}
& 
\begin{tabitemize}
\item Confidentiality
\item Integrity
\end{tabitemize}
& 
\begin{tabitemize}
\item \textit{AC-4: Information Flow Enforcement}
\item \textit{SI-10: Input Validation} 
\end{tabitemize}
\\ \cline{2-4} 
                              & 
\begin{tabitemize}
\item Employee freely inspects data in the database 
\end{tabitemize}
                              & 
\begin{tabitemize}
\item Confidentiality
\end{tabitemize}
                              & 
\begin{tabitemize}
\item \textit{AC-6: Least Privilege}  
\end{tabitemize}
\\ \hline
\end{tabular}
}
\end{table}
\vspace{-1.5em}

\subsection{Assign Effectiveness to Atomic Controls}
The analyst must now assign the effectiveness values for each identified applicable atomic control at mitigating the threats and protecting the security objectives listed in Table~\ref{tab:threatmodel}. The analyst has elected to assign qualitative ratings for the effectiveness of each atomic control that are mapped to quantitative values. The considered ratings and corresponding values are adopted and adapted from the metrics in the Common Vulnerability Scoring System (CVSS)~\cite{CVSS} and include: \textit{None} (0.0), \textit{Low} (0.2), \textit{Medium} (0.5), \textit{High} (0.8), and \textit{Very High} (0.9). No rating was assigned to the value of 1 as it is unrealistic to expect a single control to fully protect a security objective. 

With these metrics, the analyst develops the atomic payoff matrix, as illustrated in Table~\ref{tab:atomicMetricToy}. Note that none of the identified controls protect asset availability; this is fine as no threats towards availability were identified in the threat model (see Table~\ref{tab:threatmodel}). Therefore, protecting availability is not required. 

\vspace{-1em}
\begin{table}[ht]
\caption{Atomic payoff matrix for \firebird}
\label{tab:atomicMetricToy}
\centering
\small
\begin{tabular}{l|lll|lll|}
\cline{2-7}
                     & \multicolumn{3}{c|}{\textit{Database}}                           & \multicolumn{3}{c|}{\textit{User Interface}}                          \\ \cline{2-7} 
                     & \multicolumn{1}{c|}{\textit{C}}   & \multicolumn{1}{c|}{\textit{I}}   & \multicolumn{1}{c|}{\textit{A}} & \multicolumn{1}{c|}{\textit{C}}   & \multicolumn{1}{c|}{\textit{I}}   & \multicolumn{1}{c|}{\textit{A}} \\ \hline
\multicolumn{1}{|l|}{\textit{SI-10: Input Validation}} & \multicolumn{1}{l|}{Medium} & \multicolumn{1}{l|}{Very High} & None & \multicolumn{1}{l|}{Medium} & \multicolumn{1}{l|}{High} & None \\ \hline
\multicolumn{1}{|l|}{\textit{AC-3: Access Enforcement}} & \multicolumn{1}{l|}{None}   & \multicolumn{1}{l|}{None}   & None & \multicolumn{1}{l|}{Medium} & \multicolumn{1}{l|}{High} & None \\ \hline
\multicolumn{1}{|l|}{\textit{AC-4: Information Flow Enforcement}} & \multicolumn{1}{l|}{Medium} & \multicolumn{1}{l|}{Medium} & None & \multicolumn{1}{l|}{Medium} & \multicolumn{1}{l|}{Low} & None \\ \hline
\multicolumn{1}{|l|}{\textit{AC-6: Least Privilege}} & \multicolumn{1}{l|}{High} & \multicolumn{1}{l|}{None} & None & \multicolumn{1}{l|}{Medium} & \multicolumn{1}{l|}{Low} & None \\ \hline
\end{tabular}
\end{table}
\vspace{-1.5em}

\subsection{Assign Cost to Atomic Controls}
The analyst also assigns a cost for each identified atomic control as shown in Table~\ref{tab:toyCosts}. No units were used for each cost as it was decided that cost could best be represented as a unit of effort for this particular system. Additionally, the analyst's department has allocated a total budget (expressed as effort) of $B = 15$.

\vspace{-1em}
\begin{table}[ht!]
\caption{Atomic control costs for \firebird}
\label{tab:toyCosts}
\centering
\small
\begin{tabular}{|l|c|}
\hline
\multicolumn{1}{|c|}{\textbf{Control}}     & \textbf{Cost} \\ \hline
\textit{SI-10: Input Validation}            & 5    \\ \hline
\textit{AC-3: Access Enforcement}           & 6    \\ \hline
\textit{AC-4: Information Flow Enforcement} & 4    \\ \hline
\textit{AC-6: Least Privilege}              & 3    \\ \hline
\end{tabular}
\end{table}

\subsection{Specify and Generate Valid Control Combinations}
Next, the analyst must determine the valid security control combinations that could be considered for the system. To do this, they use security control algebra to specify the security control family from the mandatory and optional atomic controls identified in the previous steps. 
Recall that \textit{SI-10:~Input Validation} is a mandatory control and that \textit{AC-3:~Access Enforcement}, \textit{AC-4:~Information Flow Enforcement} and \textit{AC-6:~Least Privilege} are optional controls. 

Suppose the security analyst has determined that to implement any access enforcement policy (such as Role-Based Access Control) a least privilege approach to protecting the data in the system must first be implemented. Therefore there is a dependency between \textit{AC-3:~Access Enforcement} and \textit{AC-6:~Least Privilege}. The analyst must consider this dependency in the specification of the security control family.

Denoting the security control family as $F$, the security control family for this example is specified as the following security control algebra term and requirement relation.
\begin{eqnarray*}
F &=& \textit{SI-10}\ \odot\ \opt{\textit{AC-3},\textit{AC-4},\textit{AC-6}} \qquad
\text{such that} \qquad \requires{\textit{AC-3}}{\textit{AC-6}}{F}
\end{eqnarray*}

The possible security control combinations are generated by expanding the specification of the security control family $F$ subject to the requirement relation. The possible security control combinations for $F$ along with their costs calculated using Definition~\ref{def:costInduction} are shown in Table~\ref{tab:controlComboCosts}. Note that the security control combinations \textit{SI-10 AC-3} and \textit{SI-10 AC-3 AC-4} are not part of the security control family~$F$ because they do not respect the specified requirement relation. 

The analyst now determines the validity of the possible security control combinations according to the Budget Rule (Definition~\ref{def:rule}). 
Recall that the total budget $B$ is 15. Therefore, applying the Budget Rule for each security control combination, it is easy to see that all control combinations, except for \textit{Combo~6}, satisfy the rule and are therefore valid. As a result, \textit{Combo~6} is no longer considered.

\vspace{-1em}
\begin{table}[ht!]
\caption{Security control combination costs for \firebird}
\label{tab:controlComboCosts}
\centering
\small
\begin{tabular}{|l|l|c|}
\hline
\multicolumn{1}{|c|}{\textbf{ID}} & \multicolumn{1}{|c|}{\textbf{Security Control Combination}} & \multicolumn{1}{c|}{\textbf{Cost}} \\ \hline
\textit{Combo 1} & \textit{SI-10}                           & 5           \\ \hline
\textit{Combo 2} & \textit{SI-10 AC-4}                      & 9          \\ \hline
\textit{Combo 3} & \textit{SI-10 AC-6}                      & 8           \\ \hline
\textit{Combo 4} & \textit{SI-10 AC-3 AC-6}                 & 14          \\ \hline
\textit{Combo 5} & \textit{SI-10 AC-4 AC-6}                 & 12          \\ \hline
\textit{Combo 6} & \textit{SI-10 AC-3 AC-4 AC-6}            & 18          \\ \hline
\end{tabular}
\end{table}
\vspace{-1em}

\subsection{Construct the Game Matrix}
Now that the analyst knows all of the valid security control combinations, the game matrix can be constructed.
The payoff of each valid security control combination for each asset's security objectives is found by applying Definition~\ref{def:effInduction}. The resulting game matrix is shown in Table~\ref{tab:gameMatrixToy}. 

\begin{table}[ht!]
\caption{Game matrix for \firebird}
\label{tab:gameMatrixToy}
\centering
\small
\begin{tabular}{l|lll|lll|}
\cline{2-7}
                     & \multicolumn{3}{c|}{\textit{Database}}                           & \multicolumn{3}{c|}{\textit{User Interface}}                          \\ \cline{2-7} 
                     & \multicolumn{1}{c|}{\textit{C}}   & \multicolumn{1}{c|}{\textit{I}}   & \multicolumn{1}{c|}{\textit{A}} & \multicolumn{1}{c|}{\textit{C}}   & \multicolumn{1}{c|}{\textit{I}}   & \multicolumn{1}{c|}{\textit{A}} \\ \hline
\multicolumn{1}{|l|}{\textit{Combo 1}} & \multicolumn{1}{l|}{0.5} & \multicolumn{1}{l|}{0.9} & 0.0 & \multicolumn{1}{l|}{0.5} & \multicolumn{1}{l|}{0.8} & 0.0 \\ \hline
\multicolumn{1}{|l|}{\textit{Combo 2}} & \multicolumn{1}{l|}{0.75}   & \multicolumn{1}{l|}{0.95}   & 0.0 & \multicolumn{1}{l|}{0.75} & \multicolumn{1}{l|}{0.84} & 0.0 \\ \hline
\multicolumn{1}{|l|}{\textit{Combo 3}} & \multicolumn{1}{l|}{0.9} & \multicolumn{1}{l|}{0.9} & 0.0 & \multicolumn{1}{l|}{0.75} & \multicolumn{1}{l|}{0.84} & 0.0 \\ \hline
\multicolumn{1}{|l|}{\textit{Combo 4}} & \multicolumn{1}{l|}{0.9}    & \multicolumn{1}{l|}{0.9}    & 0.0  & \multicolumn{1}{l|}{0.875}    & \multicolumn{1}{l|}{0.968}    & 0.0  \\ \hline
\multicolumn{1}{|l|}{\textit{Combo 5}} & \multicolumn{1}{l|}{0.95} & \multicolumn{1}{l|}{0.95} & 0.0 & \multicolumn{1}{l|}{0.875} & \multicolumn{1}{l|}{0.872} &0.0 \\ \hline

\end{tabular}
\end{table}

\subsection{Play the Game}
With the game matrix constructed, the security analyst can now find a suggested security combination to protect the security objectives of the considered assets for the system. To do this, the security analyst can play the game considering different attacker profiles captured by the scenarios described below. 
Table~\ref{tab:scenarioResultsToy} presents the total effectiveness of each strategy in the game for each of the attacker objectives used in each scenario. Noteworthy effectiveness values are highlighted in bold. For strategies that have been excluded for specific attacker objectives, the corresponding effectiveness is noted as ``N/A''.\vspace{-1em}

\begin{table}[ht]
\caption{Total effectiveness of game strategies against different attacker objectives for \firebird}
\label{tab:scenarioResultsToy}
\centering
\small
\begin{tabular}{l|c|c|cc|cc|}
\cline{2-7}
	& \textbf{Scenario 1} & \textbf{Scenario 2} & \multicolumn{2}{c|}{\textbf{Scenario 3}} & \multicolumn{2}{c|}{\textbf{Scenario 4}}  \\ 
\cline{2-7} 
        & \multicolumn{1}{c|}{\textit{AO1.1}} 
	& \multicolumn{1}{c|}{\textit{AO2.1}} 
	& \multicolumn{1}{c|}{\textit{AO3.1}} & \multicolumn{1}{c|}{\textit{AO3.2}} 
	& \multicolumn{1}{c|}{\textit{AO4.1}} & \multicolumn{1}{c|}{\textit{AO4.2}}  \\
\hline
\multicolumn{1}{|l|}{\textit{Combo 1}}		& 1.0		 & 2.7		 & \multicolumn{1}{c|}{0.5}	& N/A	& \multicolumn{1}{c|}{0.9}		& N/A	\\ 
\hline
\multicolumn{1}{|l|}{\textit{Combo 2}}     	& 1.50		 & 3.29		 & \multicolumn{1}{c|}{0.75}	& N/A	& \multicolumn{1}{c|}{\textbf{0.95}}	& 1.59	\\ 
\hline
\multicolumn{1}{|l|}{\textit{Combo 3}}     	& 1.65		 & 3.39		 & \multicolumn{1}{c|}{0.75}	& N/A	& \multicolumn{1}{c|}{0.9}		& N/A	\\ 
\hline
\multicolumn{1}{|l|}{\textit{Combo 4}} 		& 1.775 	 & 3.643	 & \multicolumn{1}{c|}{\textbf{0.875}}	& \textbf{0.968}	& \multicolumn{1}{c|}{0.9}		& N/A	\\ 
\hline
\multicolumn{1}{|l|}{\textit{Combo 5}} 		& \textbf{1.825} & \textbf{3.647} & \multicolumn{1}{c|}{\textbf{0.875}}	& 0.872			& \multicolumn{1}{c|}{\textbf{0.95}}	& \textbf{1.822} \\ 
\hline
\end{tabular}
\end{table}

\textbf{Scenario 1:} This scenario considers an attacker profile where the attacker equally targets the confidentiality of the database and the confidentiality of the user interface (\textit{AO1.1}). By playing the game against this attacker, the suggested security control combination to implement is \textit{Combo~5} because it has the greatest total effectiveness (1.825) for defending against the attacker objectives. Given that all identified threats impact the confidentiality of both assets, the suggested combination includes the optional controls which maximize confidentiality across both assets. While \textit{AC-3} does not provide any protection to the confidentiality of the database, both  \textit{AC-4} and \textit{AC-6} protect confidentiality across both assets. Given that \textit{SI-10} is mandatory, \textit{Combo 5} is the logical choice. Note that by disregarding \textit{AC-3}, the threat related to "employee freely accesses and changes features provided in the interface" on the user interface is mitigated only through \textit{AC-6}. Since \textit{AC-6} is not as effective as  \textit{AC-3} for protecting integrity, the user interface's integrity is at higher risk of being violated. However, this is an acceptable risk given that the expected behaviour of the attacker is not interested in violating any integrity objectives.\\

\textbf{Scenario 2:} This scenario considers an attacker profile where the attacker equally targets all of the objectives (confidentiality, integrity, and availability) of each asset (database and user interface) (\textit{AO2.1}). By playing the game against this attacker, the suggested security control combination to implement is \textit{Combo~5} because it has the greatest total effectiveness (3.647) for defending against the attacker objectives. 
Given that this attacker profile aims to violate all security objectives on all assets, the suggested combination includes the optional controls which maximize the confidentiality and integrity across both assets. \textit{AC-4} stands out in this regard, as it effectively safeguards all security objectives unlike \textit{AC-3} and \textit{AC-6}. While \textit{AC-3} is not effective towards any of the database security objectives, \textit{AC-6} at least offers protection towards the confidentiality of the database. Given that \textit{SI-10} is mandatory, \textit{Combo 5} is again the logical choice. Note that by disregarding \textit{AC-4}, the same (acceptable) risk is imposed on the system as in Scenario 1. Also note that an assumed attacker profile targeting all objectives leads to a strategy that best balances the security objectives across all assets.\\

\textbf{Scenario 3:} This scenario considers an attacker profile where the attacker has two ordered attacker objectives to target: the confidentiality of the user interface (\textit{AO3.1}) and then  the integrity of the user interface (\textit{AO3.2}). By playing the game against this attacker, the suggested security control is determined by first considering how to best defend against the highest priority attacker objectives. This leaves \textit{Combo~4} and \textit{Combo~5} since they each have the greatest total effectiveness (0.875) for protecting the confidentiality of the user interface. Given that all controls provide the same effectiveness for the confidentiality of the user interface, any valid combination which maximizes this security objective is ideal. We now consider the next highest priority attacker objective from these possible security control combinations. Now, the suggested security control combination to implement is \textit{Combo~4} because it has a greater total effectiveness (0.968 versus 0.872 for \textit{Combo~5}) for protecting the integrity of the user interface. Given that  \textit{Combo~4} and \textit{Combo~5} differ by only one control, and that \textit{AC-3} is more effective at protecting the integrity of the user interface than \textit{AC-4}, it follows that \textit{Combo~4} is the logical choice. Note that by disregarding \textit{AC-4}, the threat related to ``commands received from unknown sources'' on the user interface is not addressed. However, this is an acceptable risk given the expected attacker profile.\\

\textbf{Scenario 4:} This scenario considers an attacker profile where the attacker has two ordered attacker objectives to target: the integrity of the database (\textit{AO4.1}) and then equally the confidentiality of the database and the integrity of the user interface (\textit{AO4.2}). By playing the game against this attacker, the suggested security control is determined by first considering how to best defend against the highest priority attacker objective. This leaves \textit{Combo~2} and \textit{Combo~5} since they each have the greatest total effectiveness (0.95) for protecting the integrity of the database. These combinations are logical as \textit{AC-4} is the only optional control that protects the integrity of the database. The next highest priority attacker objective must then be considered for these possible security control combinations. Now, the suggested security control combination to implement is \textit{Combo~5} because it has a greater total effectiveness (1.822 versus 1.59 for \textit{Combo~2}) for protecting confidentiality of the database and the integrity of the user interface. Given that \textit{Combo~5} additionally contains \textit{AC-6} which protects against the second set of attacker objectives, it follows that \textit{Combo~5} is the logical choice. Again, as \textit{Combo~5} disregards \textit{AC-4}, the same risk is imposed on the system as in Scenario 1 and Scenario 2. However, this is an acceptable risk given the expected attacker profile. \\

This illustrative example and the corresponding scenarios highlight the fact that security control selection can indeed be seen as a game, and that the suggested controls to use depends greatly on the expected attacker profiles. 
The proposed approach considers all the constraints and considerations required to perform control selection, and in contrast to existing approaches, also takes into consideration the expected attacker behaviours; an important factor to this problem that helps justify controls of interest and the risk to leave certain threats unaddressed. 

% End Section

\section{Discussion}
\label{sec:discussion}
% Begin Section

 Approaching security control selection as a game emphasizes the human element in deciding how to most effectively protect a system's assets under various considerations such as budgetary constraints. Selecting controls for a system is indeed a human-centric problem as the large number of potential controls to use from a control catalogue can be overwhelming and could lead to many mistakes in the chosen combination of security controls selected for the system. To expand on this point, selecting security controls exclusively on technical considerations while overlooking attacker behaviours is a fundamentally flawed approach in addressing this issue, as ultimately, it is humans which are conducting attacks.  Given that the proposed approach emphasizes the need to reflect on potential attacker behaviours, it prioritizes the human-centric aspect to find its solutions. Additionally, viewing this problem as a game captures the opposing dynamics of the attacker and analyst, aligning with the real-world motivations of both actors.

 Unfortunately, a limitation with many existing game-theoretic approaches from addressing cybersecurity challenges is their heavy reliance on assumptions. Specifically, game theory depends on assumptions on the players, such as the players knowing every strategy available to them, knowing the probabilities of every move, and knowing the payoff functions~\cite{OWEN2015573}. Compared to existing works, the proposed approach can be used practically as it does not rely on assigning probabilities for the likelihood of attacks succeeding, and instead focuses on general assumptions about which security objectives could be violated by the attacker. Security analysts cannot realistically predict exact probabilities of attack, but they can make informed assumptions regarding which system components might be more attractive to attackers. These considerations ensure that the proposed approach is systematic, repeatable, and realistic, thereby minimizing the influence of human bias on the results of the game and eliminating many of the required assumptions with existing game-theoretic approaches. 

While the proposed approach may seem limited by the security analyst's certainty in the effectiveness values for each atomic control, a sensitivity analysis was performed on the effectiveness metrics used in Section~\ref{sec:example} and revealed that the results of the approach were not sensitive to these values. The full details of this analysis are omitted due to space limitations. In any case, to allow the analyst to express their uncertainty, it is possible to allow them to provide multiple values when assigning the effectiveness for atomic controls. The modifications to the game under these conditions is left for future work.  
The proposed approach also currently requires manual effort on the part of the security analyst. While some tasks are unavoidably manual (\eg, selecting applicable atomic controls and assigning effectiveness and costs to those controls), the rest of the approach can be supported with automated tools. Lastly, as atomic controls are considered indivisible components, we assume costs can be independently assigned to each atomic control. From a business perspective, this assumption may not always hold as the aggregation of certain controls could result in lower total costs to implement some control combinations. We argue that this limitation is unlikely to occur unless the controls are provided by third party vendors. In such cases, the cost function outlined in Definition~\ref{def:costInduction} can be adapted to combine costs in a different manner.

% End Section

\section{Conclusions and Future Work}
\label{sec:conclusion}
% Begin Section
Ensuring effective security controls are selected for a system can greatly impact its security. In this work, a game-theoretic approach to security control selection is proposed in which a game is played by a security analyst to determine security controls which best mitigate expected attacker profiles. To create the game, the controls which are believed to secure the system must first be gathered. Following this, the effectiveness and cost of each of these controls is determined. After every possible control combination is generated, the effectiveness and cost of each combination is calculated and the game matrix can be constructed. The game can be played with many different expected attacker profiles and will suggest unique sets of controls for each. The suggested controls can help make a security analyst feel more confident in their decision to implement some controls over others. 

In future work, we aim to extend the approach to consider more than one effectiveness value for each control to account for uncertainties in the effectiveness values assigned by the analyst. This would result in the game being played with more than one game matrix. The suggested controls from these different matrices could be compared to guide a security analyst in selecting the controls for the system. Additionally, to support the calculation of the game outcomes for large systems with many controls, we aim to develop software tools to automate aspects of the approach.

% In fact, a prototype of the tool has already been developed and is available at the following link: \url{https://github.com/DylanLeveille/CSAT}.
% End Section

% Bibliography
\bibliographystyle{eptcs}
\bibliography{GandALF2024}

\begin{thebibliography}{10}
\providecommand{\bibitemdeclare}[2]{}
\providecommand{\surnamestart}{}
\providecommand{\surnameend}{}
\providecommand{\urlprefix}{Available at }
\providecommand{\url}[1]{\texttt{#1}}
\providecommand{\href}[2]{\texttt{#2}}
\providecommand{\urlalt}[2]{\href{#1}{#2}}
\providecommand{\doi}[1]{doi:\urlalt{https://doi.org/#1}{#1}}
\providecommand{\eprint}[1]{arXiv:\urlalt{https://arxiv.org/abs/#1}{#1}}
\providecommand{\bibinfo}[2]{#2}

\bibitemdeclare{article}{Almeida2018}
\bibitem{Almeida2018}
\bibinfo{author}{Lu{\'{i}}s \surnamestart Almeida\surnameend} \&
  \bibinfo{author}{Ana \surnamestart Resp{\'{i}}cio\surnameend}
  (\bibinfo{year}{2018}): \emph{\bibinfo{title}{Decision support for selecting
  information security controls}}.
\newblock {\slshape \bibinfo{journal}{Journal of Decision Systems}}
  \bibinfo{volume}{27}, pp. \bibinfo{pages}{173--180},
  \doi{10.1080/12460125.2018.1468177}.

\bibitemdeclare{article}{Bettaieb2020}
\bibitem{Bettaieb2020}
\bibinfo{author}{Seifeddine \surnamestart Bettaieb\surnameend},
  \bibinfo{author}{Seung~Yeob \surnamestart Shin\surnameend},
  \bibinfo{author}{Mehrdad \surnamestart Sabetzadeh\surnameend},
  \bibinfo{author}{Lionel~C. \surnamestart Briand\surnameend},
  \bibinfo{author}{Michael \surnamestart Garceau\surnameend} \&
  \bibinfo{author}{Antoine \surnamestart Meyers\surnameend}
  (\bibinfo{year}{2020}): \emph{\bibinfo{title}{Using machine learning to
  assist with the selection of security controls during security assessment}}.
\newblock {\slshape \bibinfo{journal}{Empirical Software Engineering}}
  \bibinfo{volume}{25}(\bibinfo{number}{4}), pp. \bibinfo{pages}{2550--2582},
  \doi{10.1007/s10664-020-09814-x}.

\bibitemdeclare{techreport}{NIST-1800-26}
\bibitem{NIST-1800-26}
\bibinfo{author}{Jennifer \surnamestart Cawthra\surnameend},
  \bibinfo{author}{Michael \surnamestart Ekstrom\surnameend},
  \bibinfo{author}{Lauren \surnamestart Lusty\surnameend},
  \bibinfo{author}{Julian \surnamestart Sexton\surnameend} \&
  \bibinfo{author}{John \surnamestart Sweetnam\surnameend}
  (\bibinfo{year}{2020}): \emph{\bibinfo{title}{Data Integrity: Detecting and
  Responding to Ransomware and Other Destructive Events}}.
\newblock \bibinfo{type}{Special Publication (NIST SP)}
  \bibinfo{number}{1800-26}, \bibinfo{institution}{{National Institute of
  Standards and Technology}}, \doi{10.6028/NIST.SP.1800-26}.

\bibitemdeclare{misc}{cisControls}
\bibitem{cisControls}
\bibinfo{author}{\surnamestart {Center for Information Security}\surnameend}
  (\bibinfo{year}{2021}): \emph{\bibinfo{title}{{CIS Critical Security Controls
  -- Version 8}}}.
\newblock \bibinfo{howpublished}{\url{https://www.cisecurity.org/controls/v8}
  [Accessed: 2024-06-21]}.

\bibitemdeclare{inproceedings}{Dewri2007}
\bibitem{Dewri2007}
\bibinfo{author}{Rinku \surnamestart Dewri\surnameend}, \bibinfo{author}{Nayot
  \surnamestart Poolsappasit\surnameend}, \bibinfo{author}{Indrajit
  \surnamestart Ray\surnameend} \& \bibinfo{author}{Darrell \surnamestart
  Whitley\surnameend} (\bibinfo{year}{2007}): \emph{\bibinfo{title}{Optimal
  security hardening using multi-objective optimization on attack tree models
  of networks}}.
\newblock In: {\slshape \bibinfo{booktitle}{Proceedings of the 14th ACM
  Conference on Computer and Communications Security}},
  \bibinfo{publisher}{Association for Computing Machinery},
  \bibinfo{address}{New York, NY, USA}, pp. \bibinfo{pages}{204--213},
  \doi{10.1145/1315245.1315272}.

\bibitemdeclare{misc}{Drake}
\bibitem{Drake}
\bibinfo{author}{Victoria \surnamestart Drake\surnameend}:
  \emph{\bibinfo{title}{{Threat Modeling}}}.
\newblock
  \bibinfo{howpublished}{\url{https://owasp.org/www-community/Threat_Modeling}
  [Accessed: 2023-12-11]}.

\bibitemdeclare{misc}{Feather2005}
\bibitem{Feather2005}
\bibinfo{author}{Martin~S. \surnamestart Feather\surnameend},
  \bibinfo{author}{Steven~L. \surnamestart Cornford\surnameend},
  \bibinfo{author}{Kenneth~A. \surnamestart Hicks\surnameend} \&
  \bibinfo{author}{Kenneth~R. \surnamestart Johnson:\surnameend}
  (\bibinfo{year}{2005}): \emph{\bibinfo{title}{Applications of tool support
  for risk-informed requirements reasoning}}.
\newblock
  \bibinfo{howpublished}{\url{https://www.researchgate.net/publication/220403935_Applications_of_tool_support_for_risk-informed_requirements_reasoning}
  [Accessed: 2024-06-21]}.

\bibitemdeclare{misc}{ITSG-33}
\bibitem{ITSG-33}
\bibinfo{author}{\surnamestart {Government of Canada}\surnameend}
  (\bibinfo{year}{2014}): \emph{\bibinfo{title}{{IT Security Risk Management: A
  Lifecycle Approach -- Security Control Catalogue}}}.
\newblock \bibinfo{howpublished}{\url{https://www.cisecurity.org/controls/v8}
  [Accessed: 2024-06-21]}.

\bibitemdeclare{article}{Hofner2011}
\bibitem{Hofner2011}
\bibinfo{author}{Peter \surnamestart H{\"{o}}fner\surnameend},
  \bibinfo{author}{Ridha \surnamestart Khedri\surnameend} \&
  \bibinfo{author}{Bernhard \surnamestart M{\"{o}}ller\surnameend}
  (\bibinfo{year}{2011}): \emph{\bibinfo{title}{An Algebra of Product
  Families}}.
\newblock {\slshape \bibinfo{journal}{Software and Systems Modeling}}
  \bibinfo{volume}{10}(\bibinfo{number}{2}), pp. \bibinfo{pages}{161--182},
  \doi{10.1007/s10270-009-0127-2}.

\bibitemdeclare{misc}{ISO-31000}
\bibitem{ISO-31000}
\bibinfo{author}{\surnamestart {International Organization for
  Standardization}\surnameend} (\bibinfo{year}{2018}):
  \emph{\bibinfo{title}{{ISO/IEC 31000:2018 Risk Management -- Guidelines}}}.
\newblock \bibinfo{howpublished}{\url{https://www.iso.org/standard/65694.html}
  [Accessed: 2024-06-21]}.

\bibitemdeclare{misc}{ISO-27002}
\bibitem{ISO-27002}
\bibinfo{author}{\surnamestart {International Organization for
  Standardization}\surnameend} (\bibinfo{year}{2022}):
  \emph{\bibinfo{title}{{ISO/IEC 27002:2022 Information security, cybersecurity
  and privacy protection -- Information security controls}}}.
\newblock \bibinfo{howpublished}{\url{https://www.iso.org/standard/75652.html}
  [Accessed: 2024-06-21]}.

\bibitemdeclare{misc}{ISO-27005}
\bibitem{ISO-27005}
\bibinfo{author}{\surnamestart {International Organization for
  Standardization}\surnameend} (\bibinfo{year}{2022}):
  \emph{\bibinfo{title}{{ISO/IEC 27005:2022 Information security, cybersecurity
  and privacy protection -- Guidance on managing information security risks}}}.
\newblock \bibinfo{howpublished}{\url{https://www.iso.org/standard/80585.html}
  [Accessed: 2023-12-11]}.

\bibitemdeclare{techreport}{NIST-RMF}
\bibitem{NIST-RMF}
\bibinfo{author}{\surnamestart {Joint Task Force Interagency Working
  Group}\surnameend} (\bibinfo{year}{2018}): \emph{\bibinfo{title}{Risk
  Management Framework for Information Systems and Organizations: A System Life
  Cycle Approach for Security and Privacy}}.
\newblock \bibinfo{type}{Special Publication (NIST SP)} \bibinfo{number}{800-37
  Revision 2}, \bibinfo{institution}{{National Institute of Standards and
  Technology}}, \doi{10.6028/NIST.SP.800-37r2}.

\bibitemdeclare{techreport}{NIST-800-53B}
\bibitem{NIST-800-53B}
\bibinfo{author}{\surnamestart {Joint Task Force Interagency Working
  Group}\surnameend} (\bibinfo{year}{2020}): \emph{\bibinfo{title}{Control
  Baselines for Information Systems and Organizations}}.
\newblock \bibinfo{type}{Special Publication (NIST SP)}
  \bibinfo{number}{800-53B}, \bibinfo{institution}{{National Institute of
  Standards and Technology}}, \doi{10.6028/nist.sp.800-53b}.

\bibitemdeclare{techreport}{NIST-800-53r5}
\bibitem{NIST-800-53r5}
\bibinfo{author}{\surnamestart {Joint Task Force Interagency Working
  Group}\surnameend} (\bibinfo{year}{2020}): \emph{\bibinfo{title}{Security and
  Privacy Controls for Information Systems and Organizations}}.
\newblock \bibinfo{type}{Special Publication (NIST SP)} \bibinfo{number}{800-53
  Revision 5}, \bibinfo{institution}{{National Institute of Standards and
  Technology}}, \doi{10.6028/NIST.SP.800-53r5}.

\bibitemdeclare{misc}{DEFEND}
\bibitem{DEFEND}
\bibinfo{author}{Peter \surnamestart Kaloroumakis\surnameend} \&
  \bibinfo{author}{Michael \surnamestart Smith\surnameend}
  (\bibinfo{year}{2020}): \emph{\bibinfo{title}{Toward a Knowledge Graph of
  Cybersecurity Countermeasures}}.
\newblock
  \bibinfo{howpublished}{\url{https://apps.dtic.mil/sti/citations/AD1156977}
  [Accessed: 2024-06-21]}.

\bibitemdeclare{article}{khalaf2021web}
\bibitem{khalaf2021web}
\bibinfo{author}{Osamah~Ibrahim \surnamestart Khalaf\surnameend},
  \bibinfo{author}{Munsif \surnamestart Sokiyna\surnameend},
  \bibinfo{author}{Youseef \surnamestart Alotaibi\surnameend},
  \bibinfo{author}{Abdulmajeed \surnamestart Alsufyani\surnameend} \&
  \bibinfo{author}{Saleh \surnamestart Alghamdi\surnameend}
  (\bibinfo{year}{2021}): \emph{\bibinfo{title}{Web Attack Detection Using the
  Input Validation Method: {DPDA} Theory}}.
\newblock {\slshape \bibinfo{journal}{Computers, Materials \& Continua}}
  \bibinfo{volume}{68}(\bibinfo{number}{3}), \doi{10.32604/cmc.2021.016099}.

\bibitemdeclare{article}{Kiesling2016}
\bibitem{Kiesling2016}
\bibinfo{author}{Elmar \surnamestart Kiesling\surnameend},
  \bibinfo{author}{Andreas \surnamestart Ekelhart\surnameend},
  \bibinfo{author}{Bernhard \surnamestart Grill\surnameend},
  \bibinfo{author}{Christine \surnamestart Strauss\surnameend} \&
  \bibinfo{author}{Christian \surnamestart Stummer\surnameend}
  (\bibinfo{year}{2016}): \emph{\bibinfo{title}{Selecting security control
  portfolios: a multi-objective simulation-optimization approach}}.
\newblock {\slshape \bibinfo{journal}{EURO Journal on Decision Processes}}
  \bibinfo{volume}{4}(\bibinfo{number}{1-2}), pp. \bibinfo{pages}{85--117},
  \doi{10.1007/s40070-016-0055-7}.

\bibitemdeclare{article}{Liu2011aa}
\bibitem{Liu2011aa}
\bibinfo{author}{Qixu \surnamestart Liu\surnameend} \& \bibinfo{author}{Yuqing
  \surnamestart Zhang\surnameend} (\bibinfo{year}{2011}):
  \emph{\bibinfo{title}{{VRSS}: A New System for Rating and Scoring
  Vulnerabilities}}.
\newblock {\slshape \bibinfo{journal}{Computer Communications}}
  \bibinfo{volume}{34}, pp. \bibinfo{pages}{264--273},
  \doi{10.1016/j.comcom.2010.04.006}.

\bibitemdeclare{techreport}{CVSS}
\bibitem{CVSS}
\bibinfo{author}{Peter \surnamestart Mell\surnameend}, \bibinfo{author}{Karen
  \surnamestart Scarfone\surnameend} \& \bibinfo{author}{Sasha \surnamestart
  Romanosky\surnameend} (\bibinfo{year}{2007}): \emph{\bibinfo{title}{The
  Common Vulnerability Scoring System {(CVSS)} and Its Applicability to Federal
  Agency Systems}}.
\newblock \bibinfo{type}{NIST Interagency Report} \bibinfo{number}{7435},
  \bibinfo{institution}{National Institute of Standards and Technology},
  \doi{10.6028/NIST.IR.7435}.

\bibitemdeclare{misc}{STRIDE}
\bibitem{STRIDE}
\bibinfo{author}{\surnamestart Microsoft\surnameend} (\bibinfo{year}{2022}):
  \emph{\bibinfo{title}{Microsoft Threat Modeling Tool -- Threats}}.
\newblock
  \bibinfo{howpublished}{\url{https://learn.microsoft.com/en-us/azure/security/develop/threat-modeling-tool-threats}
  [Accessed: 2024-06-21]}.

\bibitemdeclare{misc}{NIST-800-154}
\bibitem{NIST-800-154}
\bibinfo{author}{\surnamestart {Murugiah Souppaya and Karen
  Scarfone}\surnameend} (\bibinfo{year}{2016}): \emph{\bibinfo{title}{Guide to
  Data-Centric System Threat Modeling}}.
\newblock
  \bibinfo{howpublished}{\url{https://csrc.nist.gov/pubs/sp/800/154/ipd}
  [Accessed: 2024-06-21]}.

\bibitemdeclare{inproceedings}{Nassar2021}
\bibitem{Nassar2021}
\bibinfo{author}{Mohamed \surnamestart Nassar\surnameend},
  \bibinfo{author}{Joseph \surnamestart Khoury\surnameend},
  \bibinfo{author}{Abdelkarim \surnamestart Erradi\surnameend} \&
  \bibinfo{author}{Elias \surnamestart Bou-Harb\surnameend}
  (\bibinfo{year}{2021}): \emph{\bibinfo{title}{Game Theoretical Model for
  Cybersecurity Risk Assessment of Industrial Control Systems}}.
\newblock In: {\slshape \bibinfo{booktitle}{2021 11th IFIP International
  Conference on New Technologies, Mobility and Security (NTMS)}}, pp.
  \bibinfo{pages}{1--7}, \doi{10.1109/NTMS49979.2021.9432668}.

\bibitemdeclare{techreport}{NIST-PF}
\bibitem{NIST-PF}
\bibinfo{author}{\surnamestart {National Institute of Standards and
  Technology}\surnameend} (\bibinfo{year}{2020}): \emph{\bibinfo{title}{The
  {NIST} Privacy Framework: A Tool for Improving Privacy through Enterprise
  Risk Management}}.
\newblock \bibinfo{type}{Cybersecurity White Papers (CSWP)}
  \bibinfo{number}{10}, \bibinfo{institution}{{National Institute of Standards
  and Technology}}, \doi{10.6028/nist.cswp.10}.

\bibitemdeclare{techreport}{NIST-CSF}
\bibitem{NIST-CSF}
\bibinfo{author}{\surnamestart {National Institute of Standards and
  Technology}\surnameend} (\bibinfo{year}{2024}): \emph{\bibinfo{title}{The
  {NIST} Cybersecurity Framework {(CSF)} 2.0}}.
\newblock \bibinfo{type}{Cybersecurity White Papers (CSWP)}
  \bibinfo{number}{29}, \bibinfo{institution}{{National Institute of Standards
  and Technology}}, \doi{10.6028/NIST.CSWP.29}.

\bibitemdeclare{incollection}{OWEN2015573}
\bibitem{OWEN2015573}
\bibinfo{author}{Guillermo \surnamestart Owen\surnameend}
  (\bibinfo{year}{2015}): \emph{\bibinfo{title}{Game Theory}}.
\newblock In \bibinfo{editor}{James~D. \surnamestart Wright\surnameend},
  editor: {\slshape \bibinfo{booktitle}{International Encyclopedia of the
  Social \& Behavioral Sciences (Second Edition)}}, \bibinfo{edition}{second
  edition} edition, \bibinfo{publisher}{Elsevier}, \bibinfo{address}{Oxford},
  pp. \bibinfo{pages}{573--581}, \doi{10.1016/B978-0-08-097086-8.43045-X}.

\bibitemdeclare{article}{Park2020}
\bibitem{Park2020}
\bibinfo{author}{Jun~Young \surnamestart Park\surnameend} \&
  \bibinfo{author}{Eui~Nam \surnamestart Huh\surnameend}
  (\bibinfo{year}{2020}): \emph{\bibinfo{title}{A cost-optimization scheme
  using security vulnerability measurement for efficient security
  enhancement}}.
\newblock {\slshape \bibinfo{journal}{Journal of Information Processing
  Systems}} \bibinfo{volume}{16}(\bibinfo{number}{1}), pp.
  \bibinfo{pages}{61--82}, \doi{10.3745/JIPS.02.0128}.

\bibitemdeclare{techreport}{NIST-800-160v2r1}
\bibitem{NIST-800-160v2r1}
\bibinfo{author}{Ron \surnamestart Ross\surnameend}, \bibinfo{author}{Victoria
  \surnamestart Pillitteri\surnameend}, \bibinfo{author}{Richard \surnamestart
  Graubart\surnameend}, \bibinfo{author}{Deborah \surnamestart
  Bodeau\surnameend} \& \bibinfo{author}{Rosalie \surnamestart
  Mcquaid\surnameend} (\bibinfo{year}{2021}): \emph{\bibinfo{title}{Developing
  Cyber-Resilient Systems: A Systems Security Engineering Approach}}.
\newblock \bibinfo{type}{Special Publication (NIST SP)}
  \bibinfo{number}{800-160, Volume 2 Revision 1},
  \bibinfo{institution}{{National Institute of Standards and Technology}},
  \doi{10.6028/NIST.SP.800-160v2r1}.

\bibitemdeclare{inproceedings}{Rouland2023ab}
\bibitem{Rouland2023ab}
\bibinfo{author}{Quentin \surnamestart Rouland\surnameend},
  \bibinfo{author}{Stojanche \surnamestart Gjorcheski\surnameend} \&
  \bibinfo{author}{Jason \surnamestart Jaskolka\surnameend}
  (\bibinfo{year}{2023}): \emph{\bibinfo{title}{Eliciting a Security
  Architecture Requirements Baseline from Standards and Regulations}}.
\newblock In: {\slshape \bibinfo{booktitle}{2023 IEEE 31st International
  Requirements Engineering Conference Workshops}}, \bibinfo{series}{REW},
  \bibinfo{address}{Hannover, Germany}, pp. \bibinfo{pages}{224--229},
  \doi{10.1109/rew57809.2023.00045}.

\bibitemdeclare{article}{scholte2012}
\bibitem{scholte2012}
\bibinfo{author}{Theodoor \surnamestart Scholte\surnameend},
  \bibinfo{author}{Davide \surnamestart Balzarotti\surnameend} \&
  \bibinfo{author}{Engin \surnamestart Kirda\surnameend}
  (\bibinfo{year}{2012}): \emph{\bibinfo{title}{Have things changed now? An
  empirical study on input validation vulnerabilities in web applications}}.
\newblock {\slshape \bibinfo{journal}{Computers \& Security}}
  \bibinfo{volume}{31}(\bibinfo{number}{3}), pp. \bibinfo{pages}{344--356},
  \doi{10.1016/j.cose.2011.12.013}.

\bibitemdeclare{inproceedings}{Theodoor2012}
\bibitem{Theodoor2012}
\bibinfo{author}{Theodoor \surnamestart Scholte\surnameend},
  \bibinfo{author}{William \surnamestart Robertson\surnameend},
  \bibinfo{author}{Davide \surnamestart Balzarotti\surnameend} \&
  \bibinfo{author}{Engin \surnamestart Kirda\surnameend}
  (\bibinfo{year}{2012}): \emph{\bibinfo{title}{Preventing Input Validation
  Vulnerabilities in Web Applications through Automated Type Analysis}}.
\newblock In: {\slshape \bibinfo{booktitle}{2012 IEEE 36th Annual Computer
  Software and Applications Conference}}, pp. \bibinfo{pages}{233--243},
  \doi{10.1109/COMPSAC.2012.34}.

\bibitemdeclare{inproceedings}{Smith2017}
\bibitem{Smith2017}
\bibinfo{author}{Andrew~M. \surnamestart Smith\surnameend},
  \bibinfo{author}{Jackson~R. \surnamestart Mayo\surnameend},
  \bibinfo{author}{Vivian \surnamestart Kammler\surnameend},
  \bibinfo{author}{Robert~C. \surnamestart Armstrong\surnameend} \&
  \bibinfo{author}{Yevgeniy \surnamestart Vorobeychik\surnameend}
  (\bibinfo{year}{2017}): \emph{\bibinfo{title}{Using computational game theory
  to guide verification and security in hardware designs}}.
\newblock In: {\slshape \bibinfo{booktitle}{2017 IEEE International Symposium
  on Hardware Oriented Security and Trust (HOST)}}, pp.
  \bibinfo{pages}{110--115}, \doi{10.1109/HST.2017.7951808}.

\bibitemdeclare{book}{Straffin1993}
\bibitem{Straffin1993}
\bibinfo{author}{Philip~D. \surnamestart Straffin\surnameend}
  (\bibinfo{year}{1993}): \emph{\bibinfo{title}{Game Theory and Strategy}},
  \bibinfo{edition}{second} edition.
\newblock \bibinfo{publisher}{The Mathematical Association of America}.

\bibitemdeclare{book}{PASTA}
\bibitem{PASTA}
\bibinfo{author}{Tony \surnamestart UcedaV\'{e}lez\surnameend} \&
  \bibinfo{author}{Marco~M. \surnamestart Morana\surnameend}
  (\bibinfo{year}{2015}): \emph{\bibinfo{title}{Risk Centric Threat Modeling:
  Process for Attack Simulation and Threat Analysis}}, \bibinfo{edition}{first}
  edition.
\newblock \bibinfo{publisher}{John Wiley \& Sons}, \doi{10.1002/9781118988374}.

\bibitemdeclare{inproceedings}{Wang2010}
\bibitem{Wang2010}
\bibinfo{author}{Baoyi \surnamestart Wang\surnameend},
  \bibinfo{author}{Jianqiang \surnamestart Cai\surnameend},
  \bibinfo{author}{Shaomin \surnamestart Zhang\surnameend} \&
  \bibinfo{author}{Jun \surnamestart Li\surnameend} (\bibinfo{year}{2010}):
  \emph{\bibinfo{title}{A network security assessment model based on
  attack-defense game theory}}.
\newblock In: {\slshape \bibinfo{booktitle}{2010 International Conference on
  Computer Application and System Modeling (ICCASM 2010)}},
  \bibinfo{volume}{3}, pp. \bibinfo{pages}{V3--639--V3--643},
  \doi{10.1109/ICCASM.2010.5620536}.

\bibitemdeclare{article}{Yevseyeva2015}
\bibitem{Yevseyeva2015}
\bibinfo{author}{Iryna \surnamestart Yevseyeva\surnameend},
  \bibinfo{author}{Vitor \surnamestart Basto-Fernandes\surnameend},
  \bibinfo{author}{Michael \surnamestart Emmerich\surnameend} \&
  \bibinfo{author}{Aad \surnamestart {Van Moorsel}\surnameend}
  (\bibinfo{year}{2015}): \emph{\bibinfo{title}{Selecting Optimal Subset of
  Security Controls}}.
\newblock {\slshape \bibinfo{journal}{Procedia Computer Science}}
  \bibinfo{volume}{64}, pp. \bibinfo{pages}{1035--1042},
  \doi{10.1016/j.procs.2015.08.625}.

\end{thebibliography}

% \vfill
% \noindent\mynote{1}{\textbf{\underline{Targeted venue}}: GandALF2024
% \begin{itemize}
%     \item \textit{Abstract Deadline}: April 16, 2024 (AoE)
%     \item \textit{Paper Deadline}: April 19, 2024 (AoE)
%     \item \textit{Page Limit}: 14 pages + references and appendices (EPTCS)
%     \item \url{https://scool24.github.io/GandALF/}
% \end{itemize}
% }

\end{document}